\documentclass[amsmath,amssymb,twocolumn,amsfonts]{revtex4-2}
\usepackage{graphicx}
\usepackage{csquotes}
\usepackage[table]{xcolor}
\def \beq {\begin{equation}}
\def \eeq {\end{equation}}

\usepackage{bm}
\usepackage{bbm,mathbbol}
\usepackage{braket,bigints}
\usepackage{stmaryrd,mathtools}
\begin{document}
\title{Prospecting for lunar \textsuperscript{3}He with a radio-frequency atomic magnetometer}
\author{I. K. Kominis$^{1,2,3}$} 
\email{ikominis@uoc.gr}
\author{C. Kosmas$^4$} 
\affiliation{$^1$Department of Physics, University of Crete, Heraklion 70013, Greece}
\affiliation{$^2$School of Automation and Electrical Engineering, Zhejiang University of Science and Technology, Hangzhou 310023, China}
\affiliation{$^3$Quantum Biometronics P.C., Heraklion 71409, Greece}
\affiliation{$^4$Lunar Cargo P.C., Athens 16343, Greece}
\begin{abstract}
Mining $^3{\rm He}$ from lunar regolith has attracted significant interest in recent years due to the scarcity of $^3{\rm He}$ on Earth and its diverse applications, from cryogenics and medical imaging, to nuclear physics and future nuclear fusion. Given the stringent technical and economic challenges of mining lunar $^3{\rm He}$, precise prospecting is essential. Here we propose a prospecting methodology based on a radio-frequency atomic magnetometer, which can detect the dipolar magnetic field of thermally polarized $^3{\rm He}$ spins. With a 200 g regolith sample and an rf-magnetometer with sensitivity $1~{\rm fT/\sqrt{Hz}}$ we can detect $^3{\rm He}$ with abundance 5 ppb within a measurement time of just 5 min. The associated apparatus is lightweight and significantly more cost-effective than alternative measurement techniques. The proposed prospecting method is readily deployable and could substantially improve the technical and economic feasibility of mining lunar $^3{\rm He}$.
\end{abstract}
\maketitle 
\section{Introduction}
The Moon is not only a stepping stone for space exploration \cite{Mendel}, but also the home of valuable extraterrestrial resources \cite{Anand}. Mining lunar $^3{\rm He}$ has attracted considerable interest, on the one hand because of its limited availability on Earth, and on the other because of its potential use in nuclear fusion \cite{Kulcinski}. Unlike conventional D-T fusion, D-$^3{\rm He}$ reactions produce minimal neutron radiation, enhancing power production efficiency while minimizing long-term radioactive waste \cite{Kulcinski_2}. Its nuclear fuel potential aside, $^3{\rm He}$ is already used for low temperature physics and cryogenics \cite{cryo1,cryo2,cryo3,cryo4,cryo5}, increasingly applied to emerging quantum technologies \cite{cryo6,cryo7}, for magnetic resonance imaging \cite{mri}, and nuclear physics \cite{nuc1,nuc2}. 

Lunar regolith blankets moon's surface and contains measurable quantities of $^3{\rm He}$, at the level of 1-30 ppb \cite{Johnson,Anufriev,thesis}, implanted by the solar wind \cite{Heber,Cymes}, and measured by analyzing samples returned by the Apollo \cite{Pepin1} and later missions \cite{Pepin2,Li}. Over time, exposure to solar wind has led to the gradual accumulation of $^3{\rm He}$, particularly in titanium-rich minerals such as ilmenite \cite{Fa2010}, with an estimated quantity around $10^6$ tons \cite{Wittenberg1986}.

Extraction of $^3{\rm He}$ proceeds by heating regolith to temperatures around $1000~{\rm K}$ \cite{Schmitt2006,Song}, and separating, e.g. cryogenically, the released gases \cite{Olson}. The low abundance of $^3{\rm He}$ requires processing of large regolith mass, hence the extraction process would have to be performed on moon's surface. Given the extraction's non-trivial technical demands \cite{Matar}, it is crucial to precisely prospect for $^3{\rm He}$ and identify areas with the highest abundance. 

Here we propose a direct measurement to detect regolith-implanted $^3{\rm He}$ by use of an atomic magnetometer, in particular a radio-frequency magnetometer. The proposed method is able to detect lunar $^3{\rm He}$ from a $200~{\rm g}$ regolith sample at the level of 5 ppb within a measurement time of 5 min, while consuming minimal power, and being light-weight. Thus, the proposed detection technique is readily deployable in the lunar environment. The cost of the relevant apparatus is negligible compared to mission costs, and significantly less than alternative prospecting methodologies. Hence, from the economic perspective, the proposed technique enables swift prospecting campaigns with a compact device, potentially saving on mission duration, complexity and cost.

For completeness, we note that several authors \cite{ethos1,ethos2,ethos3,ethos4} have questioned the combined technical/economic viability of proposals for mining lunar $^3{\rm He}$, further claiming that other earth-based sources could come online, like $^3{\rm He}$-breeding reactors. While such arguments are indeed sound, we note that there are intricate, and many times surprising links between economy and technology. For example, the outlook of economic viability could change abruptly if the same infrastructure could be used for mining additional resources. In any case, we here opt to remain agnostic regarding the business case for mining lunar $^3{\rm He}$, and merely delve into the technical exercise of prospecting for this gas.

The structure of the paper is the following. In Sec. II we briefly discuss existing methodologies for detecting lunar $^3{\rm He}$. In Sec. III we present the possibility of using a radio-frequency (rf) atomic magnetometer. We conclude in Sec. IV.
\section{Existing techniques for quantifying lunar $^3{\rm He}$}
Indirect measurements are primarily based on remote sensing \cite{Nozette,Prettyman,Fa2007,Goswami,Grava,Shukla} from lunar orbiters, most of which probe for titanium, which shows strong correlation with $^3$He abundance. While remote sensing methods provide valuable first-order estimates for identifying promising regions containing $^3$He, they are inherently limited in precision and reliability as they rely on correlations rather than direct measurements. Factors like regolith depth, surface exposure history, and the efficiency of solar wind implantation introduce significant variability difficult to resolve. Thus, direct in-situ measurements remain critical for any mission aiming to mine $^3$He at economically viable scales. 

So far, such measurements rely mostly on mass spectrometers \cite{ms1,ms11,ms2,ms3,ms4,ms5,ms6,ms7}, with several variants like Quadrupole Mass Spectrometry, Resonance Ionization Mass Spectrometry, or Time-of-Flight Mass Spectrometry. Such techniques ionize atoms or molecules with different ionization schemes, and then use electromagnetic fields to separate the resulting ions based on their mass-to-charge ratios. While they require very small sample mass and provide for highly sensitive $^3{\rm He}$ detection, at the level of 1 ppb, the necessary equipment is rather bulky,  massive, and costly.
\section{Measurement with an atomic magnetometer}
\begin{figure*}
\includegraphics[width=17 cm]{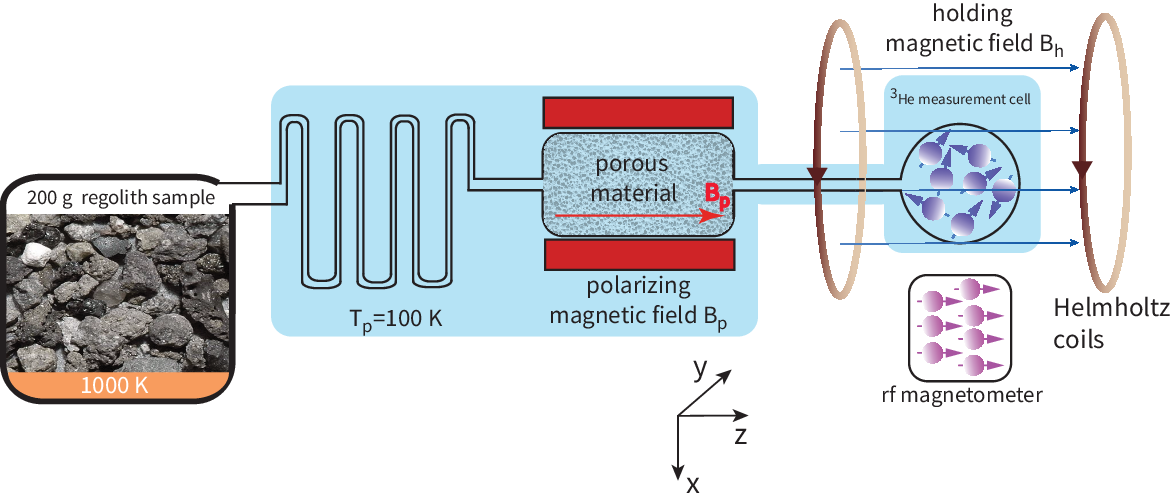}
\caption{\label{fig:schem}Schematic apparatus for measuring the abundance of $^3{\rm He}$ in regolith. Components are not drawn to scale. A regolith sample of mass $m_{\rm s}={\rm 200~g}$ is heated, and the extracted gases flow through a cold trap at $T_p=100~{\rm K}$. Then follows a polarizing magnetifc field $B_p=1~{\rm T}$, where $^3{\rm He}$ is thermally spin-polarized by flowing through a porous material designed to match diffusion time with longitudinal spin-relaxation time. Finally, the gas enters the measurement cell inside a smaller and homogeneous holding magnetic field $B_h=10~{\rm G}$. A free induction decay is induced, and the resulting oscillating magnetic dipolar field produced by $^3{\rm He}$ spins is detected by the radio-frequency magnetometer.}
\end{figure*}
Atomic magnetometers \cite{am1,am2,am3,am4,am5,am6,am7,am8,am9} detect magnetic fields by optical probing a spin-polarized alkali-metal vapor. The quantum state of the atoms in the vapor is influenced by the optical pumping and probing light, atomic collisions, internal atomic hyperfine interactions, and last but not least, by the magnetic field to be measured. 

Spin-exchange-relaxation-free-magnetometers  \cite{am10,am11} have demonstrated sub-fT magnetic sensitivity at a zero background magnetic field. On the other hand, rf-magnetometers \cite{rf1,rf2,rf3,rf4,rf5,rf6,rf7,rf8,rf9} are tuned to work at a specific non-zero bias magnetic field, and detect a weak ac magnetic field at the respective Larmor frequency. Such magnetometers utilize the fact that large spin polarization also suppresses spin-exchange relaxation, and additionally, working at high frequencies alleviates technical noise, such as magnetic noise produced by the material used to magnetically shield the former devices.

The general idea of the proposed measurement is the following. Lunar $^3{\rm He}$ shall be extracted from regolith, spin-polarized, and captured in a small measurement cell. A free-induction decay will then be induced \cite{fid}. The precessing spins of the magnetized $^3{\rm He}$ vapor will produce a dipolar magnetic field oscillating at the precession frequency. The rf magnetometer will detect this ac field, as in atomic-magnetometer-detected nuclear magnetic resonance \cite{nmr1,nmr2,nmr3,nmr4}. The envisioned measurement setup is shown in Fig. 1.
\subsection{Number of $^3{\rm He}$ atoms captured in the measurement cell}
In more detail, lunar $^3{\rm He}$ shall be extracted by heating a small regolith sample at $T_h=1000~{\rm K}$ \cite{Pepin}. Some of the released gases, like ${\rm H_2O}$, ${\rm SO_2}$ or ${\rm H_2S}$, will be captured by a cold trap at a temperature, e.g., $T_p=100~{\rm K}$. The rest of the gases not liquifying will diffuse towards the measurement volume. Such gases include $^4{\rm He}$, ${\rm H_2}$, ${\rm CO}$, ${\rm CO_2}$, ${\rm N_2}$. For this prospecting measurement, there is no need to separate them from $^3{\rm He}$, as they are nonmagnetic and their effect on $^3{\rm He}$ spin relaxation time is negligible at the low pressure of this measurement.

Now, we consider the measurement cell of volume $V_{\rm He}=1~{\rm cm}^3$, also being at temperature $T_p$. Within the temperature gradient defined by the heating temperature $T_h$ and the measurement cell temperature $T_p$, the capture efficiency ${\cal C}$ of $^3{\rm He}$ atoms by the measurement cell should approach \cite{transpiration} ${\cal C}=(V_{\rm He}/V_p)\sqrt{T_h/T_p}/(V_h/V_p+\sqrt{T_h/T_p})$, where $V_h$ and $V_p$ is the gas volume at temperature $T_h$ and $T_p$, respectively. Noting that $V_p=V_{\rm gh}+V_{\rm He}$, where $V_{\rm gh}$ is the volume of the gas handling system (also at temperature $T_p$) other than the measurement cell, and assuming $V_{\rm gh}=V_{\rm He}$ and $V_p=V_h$, it follows that ${\cal C}\approx 38\%$.

Given a regolith sample of mass $m_{\rm s}=200~{\rm g}$ and density $1.5~{\rm g/cm^3}$, and denoting by $\tilde{m}$ (with unit ppb) the $^3{\rm He}$ mass abundance, the resulting $^3{\rm He}$ atom number in the measurement cell will be
\beq
N_{\rm He}\approx 2\times 10^{16}\Big[{{\cal C}\over {38\%}}\Big]\Big[{\tilde{m}\over{1~{\rm ppb}}}\Big]\Big[{m_{\rm s}\over {200~{\rm g}}}\Big]\label{nhe}
\eeq
\subsection{Thermal spin-polarization of $^3{\rm He}$}
Before entering the measurement cell, $^3{\rm He}$ shall be pre-polarized inside a polarizing magnetic field $B_p=1~{\rm T}$ and at temperature $T_p$. Such magnetic field can be readily established with small permanent magnets in a Halbach design \cite{Halbach1,Halbach2}. In this step, $^3{\rm He}$ spins will be thermally polarized, attaining polarization $P_{\rm He}\approx \mu B_p/k_BT_p$, where $\mu=1.07\times 10^{-26}~{\rm J/T}$ is the nuclear magnetic moment of $^3{\rm He}$ and $k_B=1.38\times 10^{-23}~{\rm J/K}$ Boltzmann's constant. It follows that 
\beq
P_{\rm He}\approx 8\times10^{-6}{{\Big[B_p/1~{\rm T}\Big]}\over {\Big[T_p/100~{\rm K}\Big]}}\label{phe}
\eeq
During this pre-polarization step, $^3{\rm He}$ will be diffusing through a porous material \cite{Zhang}, which serves to slow down the diffusion of $^3{\rm He}$ atoms and thus increase the time they spend in the polarizing magnetic field. The transit time through this material should be larger than the longitudinal relaxation time $T_1$. This can be accommodated by materials like aerogels \cite{Tastevin}. For example, consider the sol-gel used to coat spin-polarized $^3{\rm He}$ glass cells \cite{Hsu}. In one example, for a 2 atm cell having 2.5 cm diameter, the longitudinal relaxation time was $T_1\approx 300~{\rm h}$.   
The $^3{\rm He}$ self-diffusion coefficient at this pressure and at room temperature is about \cite{diff} $1~{\rm cm^2/s}$, thus $^3{\rm He}$ atoms collide with the cell walls about $10^6$ times within the time $T_1$. For a porous cylinder made of the same material, having 100 nm pores, length 1 mm and diameter 1mm, $^3{\rm He}$ atoms will collide with the material about  $10^8$ times before exiting, thus there is enough time for the $^3{\rm He}$ spin polarization to equilibrate at the value $P_{\rm He}$.

The measurement cell resides in a holding magnetic field $B_h<B_p$, e.g., $B_h=10~{\rm G}$. This is because it is less straightforward to create a strong magnetic field for volumes large enough to accommodate the measurement cell. Thus we opt to pre-polarize the $^3{\rm He}$ in the \enquote{large} polarizing field $B_p$. Then the measurement can proceed in a lower, albeit homogeneous holding magnetic field $B_h$. The rf magnetometer resides in a smaller magnetic field, chose so that the Larmor resonance of the employed alkali-metal atom coincides with the precession frequency of the $^3{\rm He}$ spins.
\subsection{Measurement of the $^3{\rm He}$ dipolar magnetic field with an rf-magnetometer}
With a $\pi/2$-pulse the $^3{\rm He}$ spins can be tipped to a direction orthogonal to the holding magnetic field, and will precess about it with frequency $\omega=\gamma B_h$, where $\gamma=2\pi\times 3.24~{\rm kHz/G}$. Let $\mathbf{z}$ be the unit vector along the common direction of the polarizing magnetic field, the holding magnetic field, and the initial magnetization of $^3{\rm He}$. After the $\pi/2$-pulse, $^3{\rm He}$ spins will precess in the x-y plane (see Fig. 1), and their total magnetic moment will be $\mathbf{m}=M(\mathbf{x}\cos\omega t+\mathbf{y}\sin\omega t)e^{-t/T_2}$, where $M=\mu N_{\rm He}P_{\rm He}$, with $N_{\rm He}$ and $P_{\rm He}$ given by Eq. \eqref{nhe} and Eq. \eqref{phe}, respectively, and $T_2$ being the transverse spin relaxation time.

Assuming a spherical measurement cell, the dipolar magnetic field produced by magnetized $^3{\rm He}$ gas at a distance $R$ away from the measurement cell's center along the $x$-axis will be $B_{\rm He}e^{-t/T_2}\cos\omega t$, where the magnitude $B_{\rm He}=\mu_0M/2\pi R^3$, with $\mu_0=4\pi\times 10^{-7}~{\rm Tm/A}$ being vacuum's magnetic permeability. This magnetic field is to be sensed by the rf magnetometer. Given that the magnetometer sensor cell is not point-like, we assume an average distance between center of the sensor cell and center of the measurement cell $R\approx 3~{\rm cm}$. Then,
\beq
B_{\rm He}=\big(12~{\rm aT}\big){{\Big[B_p/1~{\rm T}\Big]}\over {\Big[T_p/100~{\rm K}\Big]}}\Big[{{\cal C}\over {38\%}}\Big]\Big[{\tilde{m}\over{1~\rm ppb}}\Big]\Big[{m_{\rm s}\over{200~{\rm g}}}\Big]
\eeq

The $^3{\rm He}$ partial pressure in the measurement cell is about 0.2 torr. We assume the rest of the gases being released from heating the regolith sample have a mass abundance similar to $^3{\rm He}$, thus we consider a total pressure of 1 torr in the measurement cell.  At such low pressures and for a realistic homogeneity of the holding magnetic field, with gradient at the level of $|\nabla B_h|\approx 1~{\rm mG/cm}$, the transverse spin relaxation time of $^3{\rm He}$ is on the order of 1 h \cite{Gemmel}. Indeed, at such temperature (100 K) and pressure (1 torr), the $^3{\rm He}$ self-diffusion coefficient is \cite{diff} $D\approx 300~{\rm cm^2/s}$. The transverse spin relaxation time is \cite{Cates} $T_2\approx 175D/16\gamma^2R^4|\nabla B_h|^2$, where $R\approx 0.6~{\rm cm}$ is the measurement cell radius for a measurement cell volume $V_{\rm He}=1~{\rm cm^3}$. Thus $T_2\approx 2400~{\rm s}$.

Currently, rf magnetometers have sensitivities around $\delta B=1.0~{\rm fT/\sqrt{Hz}}$ \cite{rf4,rf6,rf6}. Thus, for a measurement time $\tau=300~{\rm s}$ one can detect a magnetic field $B_{\rm He}$ at the level of $\delta B/\sqrt{\tau}=0.06~{\rm fT}$, which translates to measuring the abundance of regolith-implanted $^3{\rm He}$ with sensitivity 5 ppb. In other words, the sensitivity, $\delta\tilde{m}$, of the proposed measurement can be expressed as a function of all parameters as 
\begin{widetext}
\beq
{{\delta\tilde{m}}\over {5~{\rm ppb}}}={{\Big[{{\delta B/1.0~{\rm {{fT}\over\sqrt{Hz}}}}\Big]}}\over {\big(12~{\rm aT}\big)\sqrt{\Big[\tau/300~{\rm s}\Big]}}}
{{\Big[T_p/100~{\rm K}\Big]}\over {\Big[B_p/1~{\rm T}\Big]}}{1\over{\Big[m_{\rm s}/200~{\rm g}\Big]\Big[{\cal C}/38\%\Big]}}\label{abund}
\eeq
\end{widetext}
As a consistency check, the parameter dependence of $\delta\tilde{m}$ makes intuitive sense: $\delta\tilde{m}$ is reduced by reducing $\delta B$ (increasing the magnetometer sensitivity), increasing measurement time $\tau$, increasing the sample mass $m_s$ (more $^3{\rm He}$ atoms extracted), reducing the temperature or increasing the polarizing magnetic field (larger $^3{\rm He}$ spin polarization), increasing the capture efficiency (more $^3{\rm He}$ atoms). 

We note that the numerical values of the relevant parameters entering Eq. \eqref{abund} are what we think reasonable and  indicative for the workings of this measurement. In an actual realization, several technical design limitations might require different choices for those parameter values. The way the final result is expressed in Eq. \eqref{abund} can readily accommodate other choices of the parameters and easily lead to the corresponding value for $\delta\tilde{m}$ by inspection. 
\section{Discussion}
Here we wish to discuss the deployability of the proposed measurement in the lunar environment. Regarding apparatus volume, we note that there is steady progress towards developing compact atomic magnetometers \cite{Schwindt,Kitching,Stolz,Mitchell}, thus the magnetic sensor, including the associated electronics, should not contribute significantly to the volume of the apparatus of Fig. 1. The most voluminous component should be the Helmholtz coil producing the holding magnetic field. For a 5 cm radius coil, this volume is about $400~{\rm cm}^3$. Aside the regolith sample of volume somewhat larger than $100~{\rm cm}^3$, the rest of the apparatus consists of small tubing for gas handling, and thus a total volume less than 1 lt seems feasible. 

Regarding mass, the Helmholtz coil is again expected to dominate as a single component, since e.g. a 10 G field can be produced with 200 turns in total of AWG 12 wire, with a mass around 1 kg. Electronics, heating, and gas-handling systems could add another 1-2 kg, thus we expect a total mass of the apparatus smaller than 5 kg. 

Regarding power requirements, there are three major loads. The heating of the atomic magnetometer requires on the order of 100 W, and the electronics associated with the sensor around 100 W (both numbers are exaggerated on the high side). More substantial is the power requirement for the heating of the regolith material. To heat $m_{\rm s}=200~{\rm g}$ of lunar regolith to $1000~{\rm K}$, given the specific heat capacity $c\approx 1~{\rm kJ/kg/K}$, and the temperature change from 100 K of the lunar night to 1000 K, i.e. $\Delta T=900~{\rm K}$, the thermal energy required is $Q = mc\Delta T \approx 180~{\rm kJ}$. Over 300 s this translates to 600 W. We neglect the power needed to cool the parts of the apparatus required to be at low temperature, since the ambient temperature during the lunar night is as low. In total, given an available power on the order of 1 kW, a single prospecting measurement, including the heating phase and the magnetometry phase, takes about 10 min and consumes energy on the order of 100 Wh. The duration could be reduced if more power is available. Equivalently, one could increase the measurement time, thus reducing further the required sample volume according to Eq. \eqref{abund}. This point reiterates our previous comment that the specific parameter values entering Eq. \eqref{abund} will eventually relate to other mission design parameters, e.g. the available power. 

Regarding cost, it is not straightforward to estimate the cost incurred by designing and implementing space-grade hardware realizing the scheme of Fig. 1, but the required equipment altogether should cost significantly less than \$0.5 M  if it were to be used in a laboratory. 

In Table I we summarize the aforementioned performance metrics, again, not within the precision of a technical design report for a space mission, but within reasonable estimates based on current laboratory-grade technology. 

\begin{table}[hb]
\centering
\begin{tabular}{|c|c|}
\hline
\multicolumn{2}{|c|}{
    \parbox[c][4ex][c]{\linewidth}{\centering \textbf{Prospecting specs of rf-magnetometer}}
} \\
\hline
$^3{\rm He}$ Sensitivity & 5 ppb\\\hline
Regolith sample mass & $200~{\rm g}$\\\hline
Magnetometric measurement time & 5 min\\\hline
Total prospecting time & 10 min \\\hline
Equipment mass & $<5$ kg\\\hline
Equipment volume & $\sim 1$ lt\\\hline
Power & $\sim$1 kW\\\hline
Energy & $\sim$ 100 Wh\\\hline
Cost & $<$ \$500k\\\hline
\hline
\end{tabular}
\caption{Performance metrics for prospecting for lunar $^3{\rm He}$ with a radio-frequency atomic magnetometer.}
\end{table}

In summary, the proposed methodology can fit a compact and low-cost design taking advantage of the robust and simple-to-operate modern atomic magnetometers. Thus, such prospecting equipment could be readily mounted on a small lunar rover. One could imagine several exploratory prospecting campaigns performed by such a rover, which in short time could survey an area for the highest abundance of $^3{\rm He}$, before the actual mining equipment takes over.


\begin{thebibliography}{0}

\bibitem{Mendel}
W. W. Mendel, {\it Meditations on the new space vision: The moon as a stepping stone to mars}, Acta Astronautica {\bf 57}, 676 (2005).

\bibitem{Anand}
M. Anand, I. A. Crawford, M. Balat-Pichelin, S. Abanades, W. van Westrenen, G. Peraudeau, R. Jaumann, and W. Seboldt, {\it A brief review of chemical and mineralogical resources on the Moon and likely initial in situ resource utilization (ISRU) applications}, Planet. Space Sci. {\bf 74}, 42 (2012).

\bibitem{Kulcinski}
L. J. Wittenberg, J. F. Santarius, and G. L. Kulcinski, {\it Lunar source of $^3$He for commercial fusion power}, Fusion Tech. {\bf 10}, 167 (1986).

\bibitem{Kulcinski_2}
G. L. Kulcinski, G. A. Emmert, J. P. Blanchard, L. A. El-Guebaly, H. Y. Khater, J. F. Santarius, M. E. Sawan, I. N. Sviatoslavsky, L. J. Wittenberg, and R. J. Witt, {\it APOLLO - An Advanced Fuel Fusion Power Reactor for the 21st Century}, Fusion Tech., {\bf 15(2P2B)}, 1233 (1989).

\bibitem{cryo1}
H. E. Hall, {\it Helium-3 as a refrigerant}, Pure Appl. Cryogen. {\bf 6}, 363 (1966).

\bibitem{cryo2}
A. J. Leggett, A. J. (1972). {\it Interpretation of recent results on He3 below 3 mK: A new liquid phase?}, Phys. Rev. Lett. {\bf 29}, 1227 (1972).

\bibitem{cryo3}
D. D. Osheroff, R. C. Richardson, and D. M. Lee, {\it Evidence for a new phase of solid He3}, Phys. Rev. Lett. {\bf 28}, 885 (1972). 

\bibitem{cryo4}
J. C. Wheatley, {\it Experimental properties of superfluid $^3$He}, Rev. Mod. Phys. {\bf 47}, 415 (1975).

\bibitem{cryo5}
G. Frossati, {\it Experimental techniques: Methods for cooling below 300 mK}, J. Low Temp. Phys. {\bf 87}, 595 (1992).

\bibitem{cryo6}
S. Krinner, S. Storz, P. Kurpiers, P. Magnard, J. Heinsoo, R. Keller, J. L\"{u}tolf, C. Eichler, and A. Wallraff, {\it Engineering cryogenic setups for 100-qubit scale superconducting circuit systems}, EPJ Quantum Technology {\bf 6}, 2 (2019).

\bibitem{cryo7}
A. Ferraris, E. Cha, P. Mueller, K. Moselund, and C. B. Zota, {\it Cryogenic quantum computer control signal generation using high-electron-mobility transistors}, Commun. Eng. {\bf 3}, 146 (2024).

\bibitem{mri}
H. Middleton, R. D. Black, B. Saam, G. D. Cates, G. P. Cofer, R. Guenther, W. Happer, L. W. Hedlund, G. Alan Johnson, K. Juvan, and J. Swartz,  {\it MR imaging with hyperpolarized $^3$He gas}, Magn. Reson. Med. {\bf 33}, 271 (1995).

\bibitem{nuc1}
K. P. Coulter, T.E. Chupp, A. B. McDonald, C. D. Bowman, J.D. Bowman, J. J. Szymanski, V. Yuan, G. D. Cates, D. R. Benton, and E. D. Earle, {\it Neutron polarization with a polarized $^3$He spin filter}, Nucl. Instrum. Methods Phys. Res. A {\bf 288}, 463 (1990).

\bibitem{nuc2}
P. L. Anthony et al., {\it Determination of the neutron spin structure function}, Phys. Rev. Lett.  {\bf 71}, 959 (1993).

\bibitem{Johnson}
J. R. Johnson, T. D. Swindle, and P. G. Lucey, {\it Estimated solar wind-implanted helium-3 distribution on the Moon}, Geophys. Res. Lett. {\bf 26} 385 (1999).

\bibitem{Anufriev}
G. S. Anufriev, {\it Hopping diffusion of helium isotopes from samples of lunar soil}, Phys. Solid State {\bf 52}, 2058 (2010).

\bibitem{thesis}
B. O’Reilly, {\it Lunar exploration of \textsuperscript{3}He}, Undergraduate Thesis, The Ohio State University (2016).

\bibitem{Heber}
V. S. Heber, H. Baur, and R. Wieler, {\it Helium in lunar samples analyzed by high-resolution stepwise etching: Implications for the temporal constancy of solar wind isotopic composition}, Astrophys. J. {\bf 597}, 602 (2003).

\bibitem{Cymes}
B. A. Cymes, K. D. Burgess, and R. M. Stroud, {\it Helium reservoirs in iron nanoparticles on the lunar surface}. Commun. Earth Environ. {\bf 5}, 189 (2024).

\bibitem{Pepin1}
R. O. Pepin, L. E. Nyquist, D. Phinney, and D. C. Black, {Isotopic composition of rare gases in lunar samples}, Science {\bf 167}, 550 (1970).

\bibitem{Pepin2}
R. O. Pepin, D. J. Schlutter, R. H. Becker, and D. B. Reisenfeld, {\it Helium, neon, and argon composition of the solar wind as recorded in gold and other Genesis collector materials}, Geochim. Cosmochim. Acta {\bf 89}, 62 (2012).

\bibitem{Li}
A. Li et al. {\it Taking advantage of glass: capturing and retaining the helium gas on the moon}, Mater. Futures {\bf 1}, 035101 (2022).

\bibitem{Fa2010}
W. Fa and Y.-Q. Jin, {\it Global inventory of Helium-3 in lunar regoliths estimated by a multi-channel microwave radiometer on the Chang-E1 lunar satellite}, Chinese Sci. Bulletin {\bf 55}, 4005 (2010).

\bibitem{Wittenberg1986}
L. J. Wittenberg, J. F. Santarius, and G. L. Kulcinski, {\it Lunar source of $^3$He for commercial fusion power}, Fusion Tech. {\bf 10}, 167 (1986).

\bibitem{Song}
H. Song J. Zhang, Y. Sun, Y. Li, X. Zhang, D. Ma and J. Kou, {\it Theoretical study on thermal release of Helium-3 in lunar ilmenite}, Minerals {\bf 11}, 319 (2021).

\bibitem{Schmitt2006}
H. H. Schmitt, {\it Return to the Moon: Exploration, Enterprise, and Energy in the Human Settlement of Space}, Springer (2006).

\bibitem{Olson}
A. D. Olson, {\it Lunar Helium-3: mining concepts, extraction research, and potential ISRU synergies}, ASCEND 2021, doi:10.2514/6.2021-4237.

\bibitem{Matar}
S. Matar, {\it Energy analysis of extracting helium-3 from the moon}, PhD Dissertation, Plitecnico di Torino (2020).

\bibitem{ethos1}
M. Williams Pontin, {\it Mining the moon}, MIT Technology Review (2007).

\bibitem{ethos2}
I. A. Crawford, {\it Lunar resources: A review}, arXiv:1410.6865.

\bibitem{ethos3}
D. Beike, {\it Mining of helium-3 on the moon: resource, technology, and commerciality - A business perspective}, in {\it Energy Resources for Human Settlement in the Solar System and Earth’s Future in Space}, Edited by W. A. Ambrose, J. F. Reilly II, D. C. Peters, American Assocation of Petroleum Geologists (2013).

\bibitem{ethos4}
D. Day, {\it The helium-3 incantation}, The Space Review (2015).

\bibitem{Nozette}
S. Nozette et al., {\it The Clementine mission to the moon: Scientific overview}, Science {\bf 266}, 1835 (1994).

\bibitem{Prettyman}
T. H. Prettyman, J. J. Hagerty, R. C. Elphic, W. C. Feldman, D. J. Lawrence, G. W. McKinney, and D. T. Vaniman, {\it Elemental composition of the lunar surface: Analysis of gamma ray spectroscopy data from Lunar Prospector}, J. Geophys. Res. Planets {\bf 111}, E12007 (2006).

\bibitem{Fa2007}
W. Fa and Y.-Q. Jin, {\it Quantitative estimation of helium-3 spatial distribution in the lunar regolith layer}, Icarus {\bf 190}, 15 (2007).

\bibitem{Goswami}
J. N. Goswami and M. Annadurai, {\it Chandrayaan-1: India's first planetary science mission to the moon}, 40th Lunar and Planetary Science Conference (2009).

\bibitem{Grava}
C. Grava, K. D. Retherford, D. M. Hurley, P. D. Feldman, G. R. Gladstone, T. K. Greathouse, J. C. Cook, S. A. Stern, W. R. Pryor, J. S. Halekas, D. E. Kaufmann, {\it Lunar exospheric helium observations of LRO/LAMP coordinated with ARTEMIS}, Icarus {\bf 273}, 36 (2016).

\bibitem{Shukla}
S. Shukla, V. Tolpekin, S. Kumar, and A. Stein, {\it Investigating the retention of solar wind implanted Helium-3 using M3 spectroscopy and bistatic miniature radar}, Remote Sens. {\bf 12}, 3350 (2020).
\bibitem{ms1}
K. Wendt, K. Blaum, B. A. Bushaw, C. Gr\"{u}ning, R. Horn, G. Huber, J. V. Kratz, P. Kunz, P. M\"{u}ller, W. N\"{o}rtersh\"{a}user, M. Nunnemann, G. Passler, A. Schmitt, N. Trautmann, and A. Waldek, {\it Recent developments in and applications of resonance ionization mass spectrometry}, Fresenius J. Anal. Chem. {\bf 364}, 471 (1999).

\bibitem{ms11}
D. Demange M. Grivet, H. Pialot, and A. Chambaudet, {\it Indirect tritium determination by an original 3He ingrowth method using a standard helium leak detector mass spectrometer}, Anal. Chem. {\bf 74}, 3183 (2002).

\bibitem{ms2}
J. Benedikt, A. Hecimovic, D. Ellerweg and A. von Keudell, {\it Quadrupole mass spectrometry of reactive plasmas}, J. Phys. D: Applied Phys. {\bf 45}, 403001 (2012).

\bibitem{ms3}
P. R. Mahaffy et al., {\it The neutral mass spectrometer on the lunar atmosphere and dust environment explorer mission}, Space Sci. Rev. {\bf 185}, 27 (2014).

\bibitem{ms4}
L. Hofer, P. Wurz, A. Buch, M. Cabane, P. Coll, D. Coscia, M. Gerasimov, D. Lasi, A. Sapgir, C. Szopa, and M. Tulej, {\it Prototype of the gas chromatograph–mass spectrometer to investigate volatile species in the lunar soil for the Luna-Resurs mission}, Planetary Space Sci. {\bf 111}, 126 (2015).

\bibitem{ms5}
N. M. Curran, M. Nottingham, L. Alexander, I. A. Crawford, E. F\"{u}ri, and K. H. Joy, {\it A database of noble gases in lunar samples in preparation for mass spectrometry on the Moon}, Planet. Space Sci. {\bf 182}, 104823 (2020).

\bibitem{ms6}
R. Arevalo Jr, Z. Ni, and R. M. Danell, {\it Mass spectrometry and planetary exploration: A brief review and future projection}, J. Mass. Spectrom. {\bf 55}, e4454 (2020).

\bibitem{ms7}
P. Will, H. Busemann, M. E. I. Riebe, and C. Maden, {\it Indigenous noble gases in the Moon's interior}, Science Adv. {\bf 8}, eabl4920 (2022).

\bibitem{am1}
D. Budker, D. F. Kimball, S. M. Rochester, V. V. Yashchuk, and M. Zolotorev, {\it Sensitive magnetometry based on nonlinear magneto-optical rotation}, Phys. Rev. A {\bf 62}, 043403 (2000).
 
\bibitem{am2}
E. B. Aleksandrov, M. V. Balabas., A. K. Vershovskii, and A. S. Pazgalev, {\it Experimental demonstration of the sensitivity of an optically pumped quantum magnetometer}. Tech. Phys. {\bf 49}, 779 (2004).

\bibitem{am3}
S. Groeger, G. Bison, J. L. Schenker, R. Wynands, and A. Weis, {\it A high-sensitivity laser-pumped Mx magnetometer}, Eur. Phys. J. D {\bf 38}, 239 (2006).

\bibitem{am4}
D. Budker and M. V. Romalis, {\it Optical magnetometry}, Nature Phys. {\bf 3}, 227 (2007).

\bibitem{am5}
V. Shah, G. Vasilakis, and M. V. Romalis, {\it High bandwidth atomic magnetometry with continuous quantum nondemolition measurements}, Phys. Rev. Lett. {\bf 104}, 013601 (2010).

\bibitem{am6}
H. B. Dang, A. C. Maloof, and M. V. Romalis, {\it Ultra-high sensitivity magnetic field and magnetization measurements with an atomic magnetometer}, Appl. Phys. Lett. {\bf 97}, 151110 (2010).

\bibitem{am7}
J. S. Bennett, B. E. Vyhnalek, H. Greenall, E. M. Bridge, F. Gotardo, S. Forstner, G. I. Harris, F. A. Miranda, and W. P. Bowen, {\it Precision magnetometers for aerospace applications: A review}, Sensors {\bf 21}, 5568 (2021).

\bibitem{am8}
Y. Lu, T. Zhao, W. Zhu, L. Liu, X. Zhuang, G. Fang, and X. Zhang, {\it Recent progress of atomic magnetometers for geomagnetic applications}, Sensors {\bf 23}, 5318 (2023).

\bibitem{am9}
X. Bai, K. Wen, D. Peng, S. Liu, and L. Luo, {\it Atomic magnetometers and their application in industry}, Front. Phys. {\bf 11}, 1212368 (2023).

\bibitem{am10}
J. C. Allred, R. N. Lyman, T. W. Kornack, and M. V. Romalis, {\it High-sensitivity atomic magnetometer unaffected by spin-exchange relaxation}, Phys. Rev. Lett. {\bf 89}, 130801 (2002).

\bibitem{am11}
I. K. Kominis, T. W. Kornack, J. C. Allred, and M. V. Romalis, {\it A subfemtotesla multichannel atomic magnetometer}, Nature {\bf 422}, 596 (2003).

\bibitem{rf1}
S.-K. Lee, K. L. Sauer, S. J. Seltzer, O. Alem, and M. V. Romalis, {\it Subfemtotesla radio-frequency atomic magnetometer for detection of nuclear quadrupole resonance}, Appl. Phys. Lett. {\bf 89}, 214106 (2006).

\bibitem{rf2}
I. M. Savukov, S. J. Seltzer, and M. V. Romalis, {\it Detection of NMR signals with a radio-frequency atomic magnetometer}, J. Magn. Reson. {\bf 185}, 214 (2007).

\bibitem{rf3}
O. Alem, K. L. Sauer, and M. V. Romalis, {\it Spin damping in an rf atomic magnetometer}, Phys. Rev. A {\bf 87}, 013413 (2013).

\bibitem{rf4}
D. A. Keder, D. W. Prescott, A. W. Conovaloff, and K. L. Sauer, {\it An unshielded radio-frequency atomic magnetometer with sub-femtoTesla sensitivity}, AIP Advances {\bf 4}, 127159 (2014).

\bibitem{rf5}
P. Bevington, R. Gartman, D. J. Botelho, R. Crawford, M. Packer, T. M. Fromhold, and W. Chalupczak, {\it Object surveillance with radio-frequency atomic magnetometers}, Rev. Sci. Instrum. {\bf 91}, 055002 (2020).

\bibitem{rf6}
P. Bevington and W. Chalupczak, {\it Different configurations of radio-frequency atomic magnetometers - A comparative study}, Sensors {\bf 22}, 9785253 (2022).

\bibitem{rf7}
C.Z. Motamedi and K.L. Sauer, {\it Magnetic Jones vector detection with rf atomic magnetometers}, Phys. Rev. Appl. {\bf 20}, 014006 (2023).

\bibitem{rf8}
D. J. Heilman, K. L. Sauer, D. W. Prescott, C. Z. Motamedi, N. Dural, M. V. Romalis, and T. W. Kornack, {\it Large-scale multipass two-chamber rf atomic magnetometer}, Phys. Rev. Appl. {\bf 22}, 054024 (2024).

\bibitem{rf9}
W. Xiao, X. Liu, T. Wu, X. Peng, and H. Guo, {\it Radio-frequency magnetometry based on parametric resonances}, Phys. Rev. Lett. {\bf 133}, 093201 (2024).

\bibitem{fid}
M. Levitt, {\it Spin dynamics: Basics of nuclear magnetic resonance} (John Wiley \& Sons, New York, 2008).

\bibitem{nmr1}
I. M. Savukov and M. V. Romalis, {\it NMR detection with an atomic magnetometer}, Phys. Rev. Lett. {\bf 94}, 123001 (2005).

\bibitem{nmr2}
M. P. Ledbetter, I. M. Savukov, D. Budker, V. Shah, S. Knappe, J. Kitching, D. J. Michalak, S. Xu, and A. Pines, {\it Zero-field remote detection of NMR with a microfabricated atomic magnetometer}, Proc. Natl. Acad. Sci. U.S.A. {\bf 105}, 2286 (2008).

\bibitem{nmr3}
G. Liu, X. Li, X. Sun, J. Feng, C. Ye, and X. Zhou, {\it Ultralow field NMR spectrometer with an atomic magnetometer near room temperature}, J. Magn. Res. {\bf 237}, 158 (2013).

\bibitem{nmr4}
D. A. Barskiy, J. W. Blanchard, D. Budker, J. Eills, S. Pustelny, K. F. Sheberstov, M. C. D. Tayler, and A. H. Trabesinger, {\it Zero- to ultralow-field nuclear magnetic resonance}, Prog. Nucl. Magn. Reson. Spectrosc. {\bf 148-149},101558 (2025).

\bibitem{Pepin}
R. O. Pepin, L. E. Nyquist, D. Phinney, and D. C. Black, {\it Rare gases in Apollo 11 lunar material}, Proc. Apollo 11 Lunar Sci. Conf. {\bf 2}, 1435 (1970).

\bibitem{Zhang}
L. Zhang, K. Wu, Z. Chen, X. Yu, J. Li, S. Yang, G. Hui, and M. Yang. {\it Gas storage and transport in porous media: From shale gas to helium-3}, Planetary Space Sci. {\bf 204}, 105283 (2021).

\bibitem{Halbach1}
K. Halbach, {\it Design of permanent multipole magnets with oriented rare earth cobalt material}, Nuclear Instruments and Methods {\bf 169}, 1 (1980).

\bibitem{Halbach2}
H. Raich and P. Bl\"{u}mler, {\it Design and construction of a dipolar Halbach array with a homogeneous field from identical bar magnets: NMR Mandhalas}, Concepts Magn. Reson. {\bf 23B}, 16 (2004).

\bibitem{Tastevin}
G. Tastevin and P.-J. Nacher, {\it NMR measurements of hyperpolarized 3He gas diffusion in high porosity silica aerogels}, J. Chem, Phys. {\bf 123}, 064506 (2005).

\bibitem{Hsu}
M. F. Hsu, G. D. Cates, I. Kominis, I. A. Aksay, and D. M. Dabbs, {\it Sol-gel coated glass cells for spin-exchange polarized $^3$He}, Appl. Phys. Lett. {\bf 77}, 2069 (2000).

\bibitem{diff}
D. R. Burgess Jr, {\it Self-diffusion and binary-diffusion coefficients in gases}, NIST Technical Note 2279 (2024).

\bibitem{transpiration}
Y. Wu, {\it Theory of thermal transpiration in a Knudsen gas}, J. Chem. Phys. {\bf 48}, 889 (1968).

\bibitem{Gemmel}
C. Gemmel, W. Heil, S. Karpuk, K. Lenz, Ch. Ludwig, Yu. Sobolev, K. Tullney, M. Burghoff, W. Kilian, S. Knappe-Gr\"{u}neberg, W. M\"{u}ller, A. Schnabel, F. Seifert, L. Trahms, and St. Baeßler, {\it Ultra-sensitive magnetometry based on free precession of nuclear spins}, Eur. Phys. J. D {\bf 57}, 303 (2010).

\bibitem{Cates}
G. D. Cates, S. R. Schaefer, and W. Happer, {\it Relaxation of spins due to field inhomogeneities in gaseous samples at low magnetic fields and low pressures}, Phys. Rev. A {\bf 37}, 2877 (1988).

\bibitem{Schwindt}
P. D. D. Schwindt, S. Knappe, V. Shah, L. Hollberg, J. Kitching, L.-A. Liew and J. Moreland, {\it Chip-scale atomic magnetometer}, Appl. Phys. Lett. {\bf 85}, 6409 (2004).

\bibitem{Kitching}
J. Kitching, {\it Chip-scale atomic devices}, Appl. Phys. Rev. {\bf 5}, 031302 (2018).

\bibitem{Stolz}
G. Oelsner, R. IJsselsteijn, T. Scholtes, A. Kr\"{u}ger, V. Schultze, G. Seyﬀert, G. Werner, M. J\"{a}ger, A. Chwala, and R. Stolz, {\it Integrated optically pumped magnetometer for measurements within Earth’s magnetic field}, Phys. Rev. Applied {\bf 17}, 024034 (2022).

\bibitem{Mitchell}
H. Raghavan, M. C.D. Tayler, K. Mouloudakis, R. Rae, S. L\"{a}hteenm\"{a}ki, R. Zetter, P. Laine, J. Haesler, L. Balet, T. Overstolz, S. Karlen, and M. W. Mitchell, {\it Functionalized millimeter-scale vapor cells for alkali-metal spectroscopy and magnetometry}, Phys. Rev. Applied {\bf 22}, 044011 (2024).


\end{thebibliography}
\end{document}